\documentclass[%
 reprint,
groupedaddress,
showpacs,
showkeys,
preprintnumbers,
nobibnotes,
 amsmath,amssymb,
 aps,
prb,
]{revtex4-1}

\usepackage{IEEEtrantools}
\usepackage{array}
\usepackage{amsmath,  amsfonts, amssymb}
\usepackage{graphicx}
\usepackage{dcolumn}
\usepackage{bm}

\begin{document}

\preprint{}

\title{Localization parameters for two interacting particles in disordered two-dimensional lattices}

\author{Tirthaprasad  Chattaraj}
\email{tirtha@chem.ubc.ca}
\affiliation{%
University of British Columbia\\
Vancouver, B.C., V6T 1Z1, Canada
}%

\date{\today}

\begin{abstract}
In two dimensional disordered lattices, presence of interaction makes particles less localized than the non-interacting ones within the range of disorder strength $W \le 4$ and interaction strength $V \le 4$. If the interaction strength is higher, then particles localize more. Although, a localization-delocalization transition is not found, a transition with changes in the dominant correlations is observed. The nature of correlations between the particles as nearest neighbors  become dominant beyond certain disorder strengths.

\end{abstract}


\maketitle


\section{\label{sec:level1}Introduction \protect\\  \lowercase{}}

The case of localization of two interacting particles in one dimensional lattices has been investigated thoroughly \cite{romer, frahm, oppen, flach, pichard1, pichard3, tirtha1} since last nineties. These studies were motivated by the observation of persistent currents in 1D wires \cite{ambegaokar, levy, chandrasekhar}. Similar experimental observation of localization-delocalization transition in two-dimensional systems \cite{kravchenko, kravchenkos} had drawn attention for theoretical research on effect of interaction on such transition \cite{ortuno,vasseur, pichard}.  However, the study of localization of two interacting particles in 2D lattices brings computational difficulty with it. In a recently developed algorithm, \cite{berciu, ping, tirtha2} based on recursive calculation of two-particle Green's functions, such difficulties have been reduced and such calculations have been made possible to perform within reasonable amout of resources. Although, even after algorithmic reduction of difficulties, the task still remains challenging for large lattices. Therefore certain approximations must be employed which has been justified in detail in a separate work \cite{tirtha2}.  However, such calculations not only helps in gaining understanding on the effect of interaction on localization, but also provide information on the correlations of the particles. These correlations between particles calculated from two-particle Green's functions, reveal further details on the phases of localization in the disordered systems.

For the calculation of localization parameters, one of the prescribed approach is to calculate two-particle Green's propagator from the center of the lattice to boundaries \cite{song} with combination of scaling arguments \cite{mackinnon}. This approach has been implemented in previous studies \cite{ortuno}. However, computational difficulty limits such calculations to small system sizes which often entail significant finite size effects. The drastic approximations involved in such studies also render it difficult to make any conclusive statement on localization-delocalization transition in 2D disordered systems.

In this article, inverse participation ratio (IPR) is calculated as a localization parameter. Calculation of IPR involves computation of all Green's propagators for given interaction strength between the particles and strength of disorder of the lattice. The IPR can be taken as a macroscopic parameter dependent on the distribution of particles on the whole lattice and is less susceptible to finite size effects. Thus the IPR numbers are expected to provide a better understanding on the length scale of localization of interacting particles in finite 2D disordered systems. Additional informations on the localization are obtained from the correlations  between particles calculated from two-particle Green's functions. The correlation parameters provide more information on the underlined structure of the distribution of the particles in localized systems. 

The IPR parameters are calculated for a broad range of disorder strengths of the lattices and interaction strengths between the particles. This broad range of calculations help in gaining further insights into the interplay of disorder and interaction in 2D systems. With the additional informations on correlations, separate phases can be recognized between localized states in finite disordered 2D systems.

\section{\label{sec:level1}Method \protect\\  \lowercase{}}

In this article. the Hubbard Hamiltonian with nearest neighbor hopping and nearest neighbor interaction is considered for the study. The particles are taken as hard core bosons. This is the most general form of Hamiltonian for studies on localization in both one- and two-dimensional disordred systems.

\begin{eqnarray}\label{ham2D}
H =  \sum_{ij} \epsilon_{ij} a_{i j}^\dagger a_{i j} +  \sum_{ij} ( a_{i j}^\dagger a_{i+1 j} + a_{i j}^\dagger a_{i j+1} ) \nonumber \\
 + V \sum_{ ij} (n_{ij} n_{i+1 j} + n_{ij} n_{i j+1})
\end{eqnarray} 
Here $i, j$ are the site indices on two axes of the lattice. The onsite energy $\epsilon_{ij}$ is chosen randomly from a uniform box distribution  [$-\frac{W}{2} , \frac{W}{2}$] of width $W$, which defines the disorder strength.
The computational algorithm involved in the calculations is explained in detail in a separate article \cite{tirtha2}. The Green's elements for a given disorder strength from an initial occupation of particles at adjacent sites in the middle of the lattice are  calculated at very large times ($\tau  = 1000$) using the algorithm. The lattice was considered to have 20 sites per dimension. The approximation of maximum relative distance ($r$) was applied with $r=5$ for significant enhancement of the efficiency of the calculations without significant errors \cite{tirtha2}. The Green's elements were thus computed at sufficiently large times than required for the spreading of the particles to the lattice boundaries.

   \begin{equation}\label{dyn}
     G(i,j,\tau)  = \sum_{\omega} e^{-i \omega \tau}  G(i, j, \omega) 
    \end{equation}
Once all such Green's elements were found, that is for every site indices ($i, j$) with $|i-j| \le r$ for two particles, the joint density distribution ($\rho$), density distribution ($\varrho$) and inverse participation ratio ($\mathcal{I}$) were calculated for each realization of disorder, which were averaged afterwords over many realizations.

   \begin{equation}\label{density}
     \rho (i,j,\tau)  =  |G(i, j, \tau)|^2
    \end{equation}

   \begin{equation}\label{density}
     \varrho (i,\tau)  =  \frac{1}{2} \sum_{j\neq i}  \rho (i,j,\tau)
    \end{equation}

   \begin{equation}\label{ipr}
       \mathcal{I} =  \frac{\sum_{i}\varrho (i,\tau)^2}{\sum_{i}\varrho (i,\tau)}
    \end{equation}
For a measure of total correlation ($\zeta$) with respect to the distance between particles, the minimum step distance between the particles on the lattice ($r_s$, hamming distance) was taken as a measure of distance and the correlation elements were summed based on such distance irrespective of the locations of the particles on the lattice. 

   \begin{equation}\label{cor}
       \zeta (r_s)=  \sum_{|i-j|=r_s}\rho (i,j, \tau)
    \end{equation}
This parameter then reflects the total correlation within the density distribution in the disordered 2D lattice.

\section{\label{sec:level1}Results \protect\\  \lowercase{}}

\begin{figure}[h]
 \includegraphics[width=0.5\textwidth]{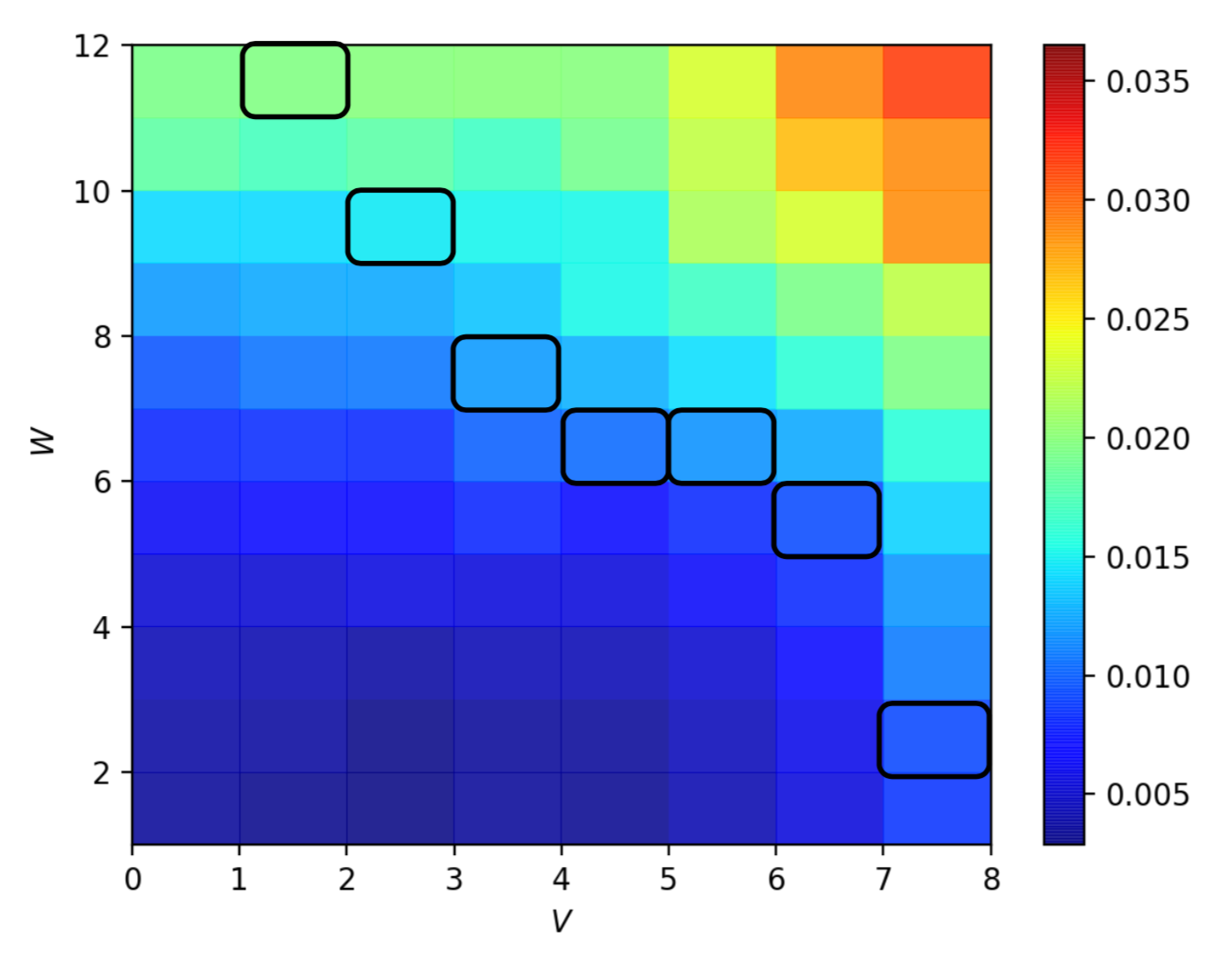};   
\caption{Two dimensional disorder-interaction diagram with black squares showing the regions where nature of correlations change. The inverse participation ratios plotted are averaged over 320 realizations of disorder.}
\label{ipr2d}
\end{figure}

The results of the calculations for medium sized finite disordered 2D lattices are shown in Fig. \ref{ipr2d}. The calculations involved combinations of disorder ($W$) and interaction strengths ($V$) within the range $1 \le W \le 12$ and $0 \le V \le 8$. The IPR numbers were found to be ranging from $0.004$ to $0.04$ within this broad range of disorder and interaction strengths. Higher IPR represent more localization of the particles. 

\begin{figure}[h]
\includegraphics[width = 0.5\textwidth]{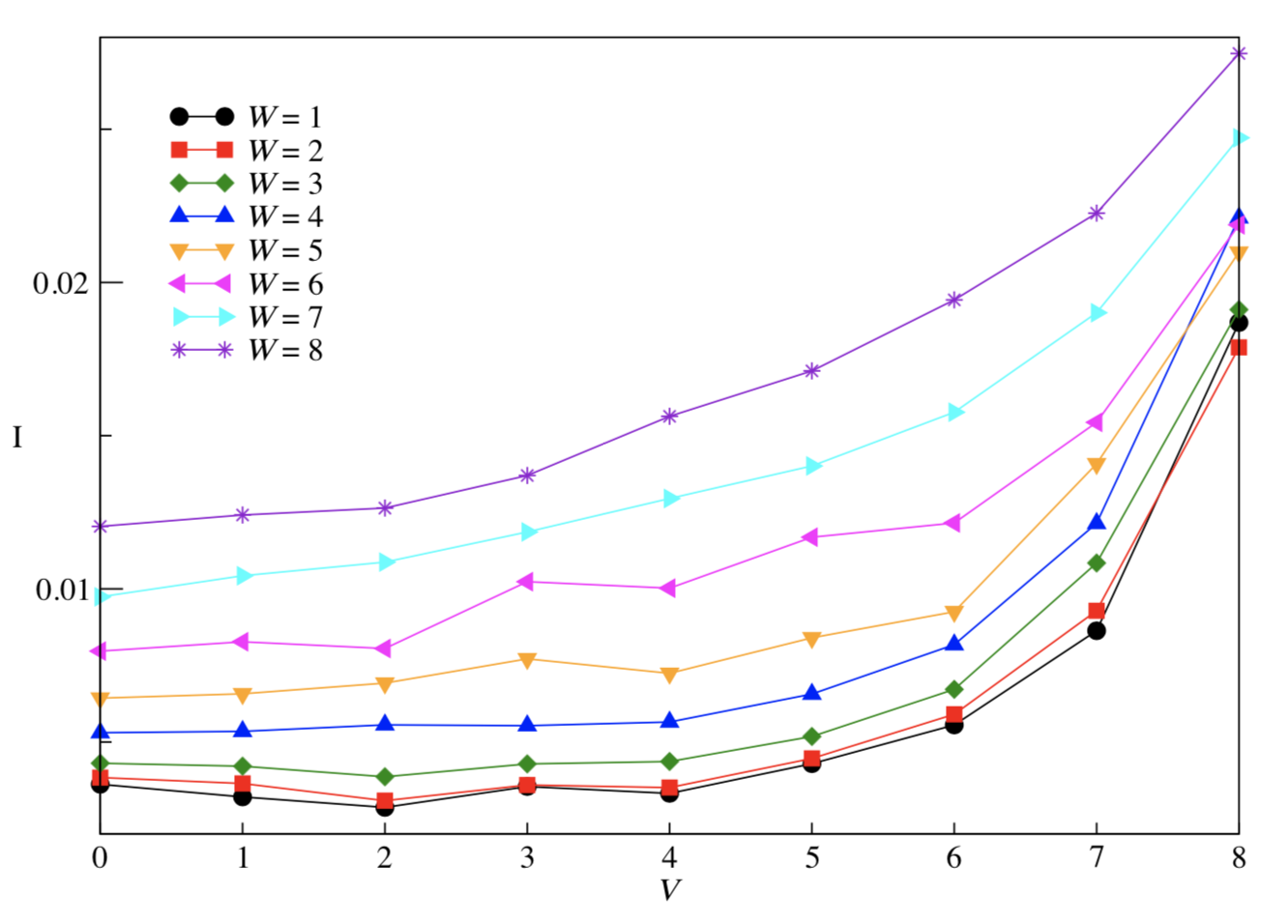}
\caption{Cross sections from the Fig. \ref{ipr2d} on disorder axis. It shows decrease in IPR from that of non-interacting particles within the range $1 \le W \le 4$ and $0 \le V \le 4$.}
\label{Vipr}
\end{figure}

Figure \ref{ipr2d} shows a reduction in localization with increase in interaction upto $V \le 4$, when compared to $V = 0$ case, for the range of disorder strength $1 \le W \le 4$. This range can be termed as weak interaction and weak disorder region, where the interacting particles have smaller IPR and lesser localization compared to non-interacting ones.  This is made clearly visible  in Fig. \ref{Vipr} with IPR numbers for fixed disorder cross sections plotted as a function of interaction strength.

Beyond the weak disorder - weak interaction regime, enhancement in localization is observed in presence of interaction.

The marked squares on Fig. \ref{ipr2d} indicates a change in the correlations between the particles in their localization. The measure of total correlation  $\zeta (r_s)$ for $r_s = 2$ is larger than that of $r_s = 1$ toward the lower IPR region of these squares.  $\zeta (1)$ becomes larger than  $\zeta (2)$ when either disorder or interaction is increased beyond these marked squares.  The following Fig. \ref{cor} exhibits one of such squares for $V = 4$, where it can be observed that  $\zeta (2) >  \zeta (1)$ beyond $W = 6$. This change in the underlined correlations between the particles in their localization can be an observable for measurements involving two-particle correlators.   This change also signifies not only a transition from low IPR to high IPR region, but also a type of underlined phase of the particles which cannot be observed from single particle density measurements.

\begin{figure}[h]
\includegraphics[width = 0.5\textwidth]{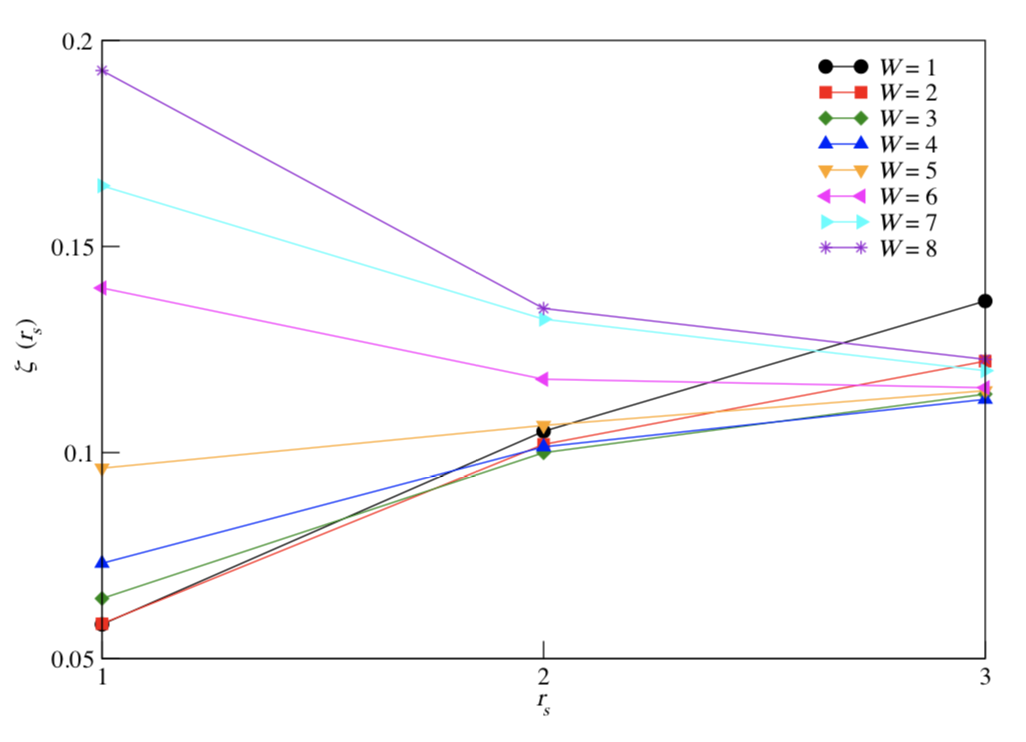}
\caption{Correlation of the two particles depending on the (hamming) distance between them irrespective of their location at lattice for $V = 4$.}
\label{cor} 
\end{figure}

The density distribution of one of such points from Fig. \ref{ipr2d} ($W=1, V=4$) is shown in Fig. \ref{density}. As the distribution shows, after averaging over 250 realizations, the final distribution appears to be localized. This  point from the Fig. \ref{ipr2d} has low IPR compared to other points of the diagram. However, the density distribution appears to be localized for a finite sized 2D disordered lattice. 

\begin{figure}[h]
\includegraphics[width = 0.5\textwidth]{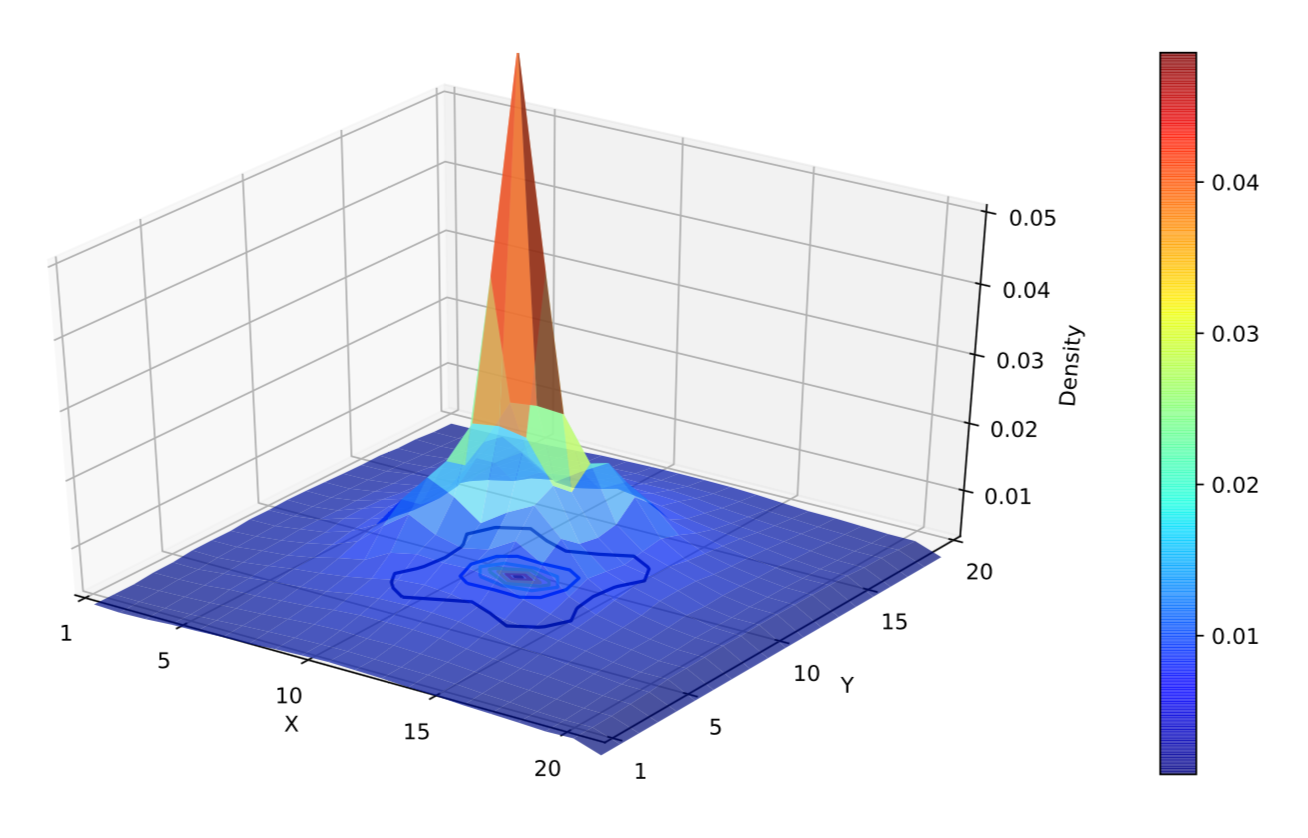}
\caption{Density distribution on the two dimensional lattice for $V=4$ and $W=1$. Averaged over 320 realizations of disorder.}
\label{density}
\end{figure}
\begin{figure}[h]
\includegraphics[width = 0.5\textwidth]{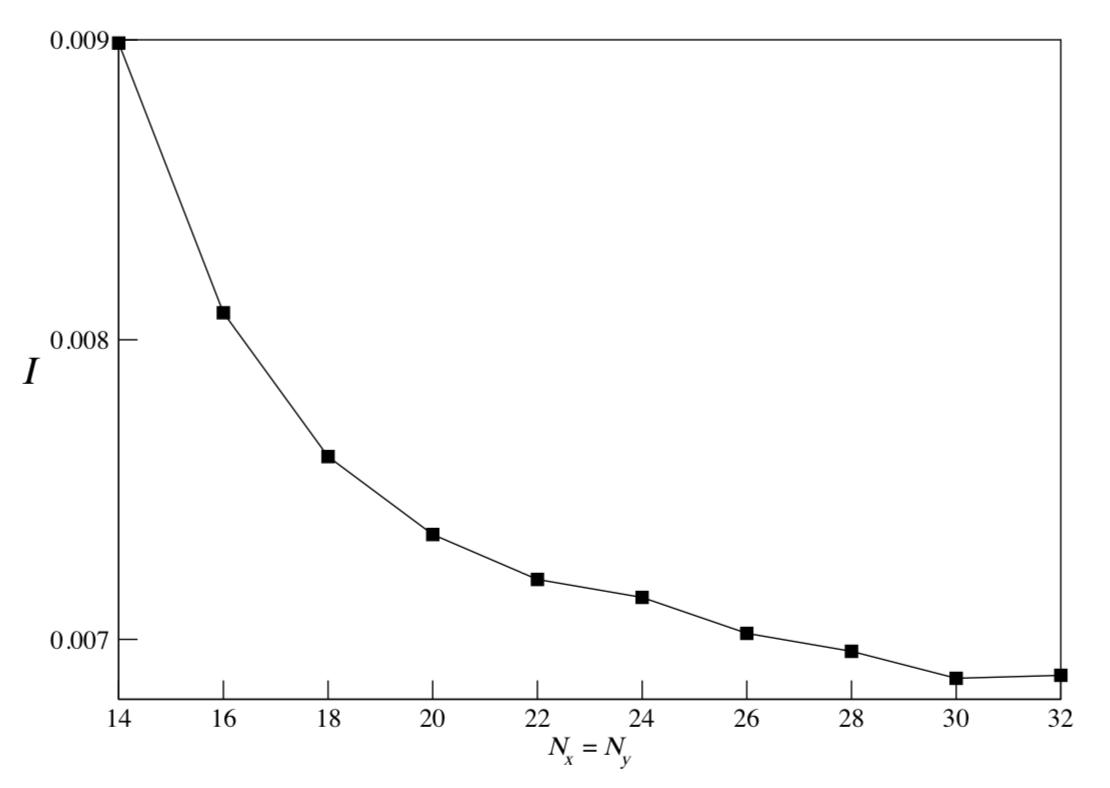}
\caption{Scaling of IPR with the lattice size for $V=4$ and $W=1$. Averaged over 50 realizations of disorder.}
\label{scaling}
\end{figure}

The scaling analysis of this specific point ($W=1, V=4$) from Fig. \ref{ipr2d} is described in Fig. \ref{scaling}, which indicates a localized state as the IPR approaches a constancy for larger system sizes. However, a conclusion on the absence of delocalization for the model under consideration at this weak interaction - weak disorder regime may not be drawn as the calculations involved the previously mentioned approximations. Any such conclusion will require analysis of the full problem which will involve further computational challenges.

The effect of increasing localization, when both interaction and disorder is strong,  can be interpreted as the effect from participation of bound state in the underlined dynamics and disorder enhancement of binding. This is also implied from the correlations between the particles. However when the bound state interpreted as the particles correlated as nearest neighbor ones, Fig. \ref{ipr2d} do not map directly with that of crossing points where such correlations change. The effect of different types of correlations in the underlined distribution might show different features of localization.

\begin{figure}[h]
\includegraphics[width=0.5\textwidth]{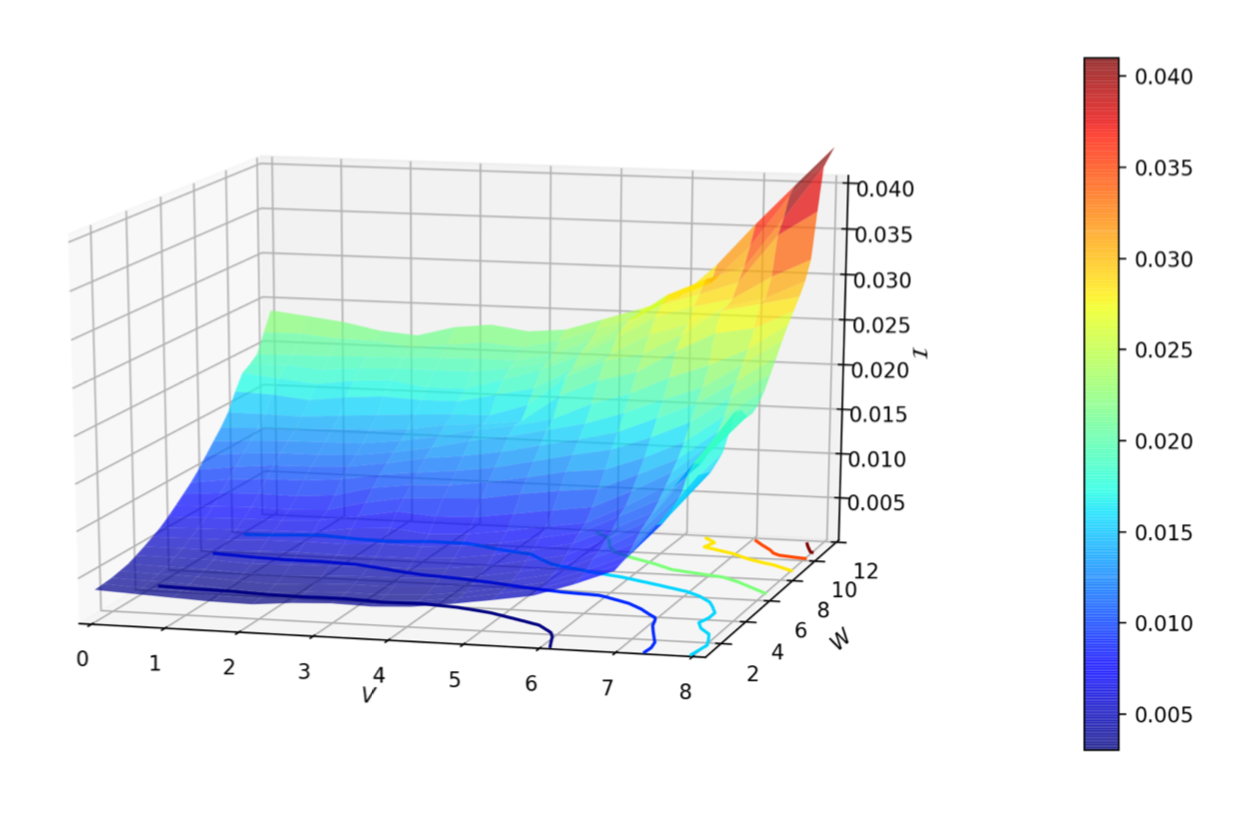}; 
\caption{Three dimensional disorder-interaction diagram. The different character of the light blue and light green contours indicate a possible difference between the two regions. The inverse participation ratios plotted are averaged over 320 realizations of disorder.}
\label{ipr3d}
\end{figure}

Accompanied with the understanding on correlations within the localized distribution of particles, one can attempt to draw separate phases of localizations. A 3D distribution of the same Fig. \ref{ipr2d} is shown in Fig. \ref{ipr3d}. The contours drawn on the lower surface of Fig. \ref{ipr3d} indicate such differences between localized phases.  While the contours with IPR $\le 0.01$ reveals a centricity towards the origin, the contours with IPR $\ge 0.02$ have centricity away from the origin on the diagram.  This can be based on the correlations between the particles. This region with IPR $\le 0.01$ shows a possibility delocalized behavior that may be corroborated with other works \cite{ortuno}. However, this study do not find any sign of delocalized behavior which may be a result of finite sized systems under consideration combined with the approximations involved int the calculations.

\section{\label{sec:level1}Conclusion \protect\\  \lowercase{}}

This study has attempted to calculate localization parameters for disordered 2D lattice systems. The calculations not only provide an understanding on the length scales of the localized interacting particles, but also present an understanding on the correlations between the particles in their localized states. The inverse participation ratios calculated for a vast range of disorder and interaction parameters reveal the effects of both interaction and disorder on localization of particles in disordered 2D systems. Although, a localization-delocalization transition is not found in this study, it doesn't exclude such possibilities. Understanding of the correlations suggest different phases of localization within the broad ranges of disorder and interaction strengths. Whether such phases indicate a localization-delocalization transition will require further study of the systems.

\nocite{*}

\bibliography{rev}

\end{document}